\documentstyle[12pt]{article}

\author{A.~T.~Banin, N.~G.~Pletn'ev\thanks{e-mail: pletnev@math.nsk.su}}
\title{Coset realizations of (super)twistor spaces and structure of
(super)twistor correspondence
}
\date{\it
Institute of Mathematics, Novosibirsk, \\
Universitetsky prosp. 4, 630090, Russia
}
\begin{document}

\begin{titlepage}
\maketitle
\vspace{2truecm}
\begin{abstract}
New types "extended" (super)conformal algebras
$G^{(\frac n2)}$ are presented. (Su\-per)twistor spaces $T$ are
subspaces in cosets $G^{(\frac n2)}/H$. The (super)twistor
correspondence has a cleary defined geometrical meaning.
\end{abstract}
\thispagestyle{empty}
\end{titlepage}

\section{Introduction}

\noindent
Due to impressing applications of complex-analytic constructions to a number
of physics problems, twistor theme makes an integral part of various
disciplines. Famous papers \cite{Penr} generated significant interest to
this theme. Firstly, it was pointed out that equations of motion for
massless fields on Minkowsky space $M^4$ with arbitrary spin could be
rewritten as Cauchy -- Riemann conditions for fields on twistor space $T$.
In this way the problem of description of fields on $M^4$ comes to
constructing of the space of complex structures on an oriented Riemann
manifold and to the transformation of the original fields to the objects of
complex algebraic geometry (fields on twistor space). It would like to apply
twistor methods for the description of gravity as an essentially nonlinear
theory. In the context of the twistor program the correspondence between
conformal classes of auto-dual solutions of Einstein equations and
deformations of functions in $CP^3$ was found in ref. \cite{Atia}. There is
analogous correspondence between the self-dual solutions of Yang-Mills (YM)
equations and the two-dimensional fiber bundle over domains $CP^3$ \cite
{Ward}, which leads to the classes of instanton and monopole solutions. This
isomorphism is a mathematical design of the physics idea that the connection
(as a dynamical field) shouldn't be defined on space-time points but on the
ways (complex null-line), which are natural for the conformal invariant and
holomorphic theories.

An additional interest in the selfduality connects with a proposition that
all integrable systems can be gotten by the dimensional reduction from $4D$
self-dual conformal invariant theories \cite{Maso}. The complexification of
Minkowsky space is the common procedure for twistor methods. The inverse
procedure --- the realification of complex Minkowsky space is done by a
sequential violation of $SL(4,C)$ symmetry. This procedure is connected with
a choice of infinitely far light cone, a real structure and a scale.

Recently, it was arisen a tendency to use twistors for the solutions of
basic problems of prepotential formulations of SUSY theories. An example,
where SUSY and twistors are necessary for each other, is the conditions of
integrability of wave equations on the light-like paths of superparticle
interacting with gauge super YM fields and supergravity. These conditions
define both the constrains and the equations of motion for SUSY formulations
of such theories \cite{Witt}.

New ''twistor-like'' reformulations of the action principle for
(super)particles and (super)strings \cite{Ferb, Soro} turned out to be
productive, because Cartan -- Penrose conditions aren't postulated but
arisen as the solutions of dynamical constrains and equations of motion. As
this take place, Siegel $k$-symmetry \cite{Sieg} is a manifestation of local
proper time SUSY, where the Grassmannian coordinate of ''target'' superspace
and the momentum component of the twistor set up one supermuliplet. Number
of the first-kind constrains in these models is sufficient to fix unphysical
degrees of freedom. To get closed algebra ''off--shell'', a covariant form
of the light-cone gauge is required. This gauge can be obtained through the
use of some auxiliary variables which parameterize coset $SL\left(
2,C\right) /H$, where $H$ is some subgroup of $SL\left( 2,C\right) $ \cite
{Soka}.

It is well known, that the problems associated with the applying of the
twistor approach to massive particles are arisen from the absence of
conformal invariance of the action. Nevertheless, the using of holomorphic
functions of several twistor variable allows us to find the twistor version
of d'Alambert operator on twistor space. The simple calculation of the
degrees of freedom shows that the twistor method has some excess parameters,
which can be associated with some gauge symmetry. It is well known that
gauge invariance determines the main structure of any gauge theory. For one
example, ever the theory of massless particle has $U(1)$ chiral symmetry
that leaves kinematic twistors to be invariant.

According to Penrose, the natural symplectic structure of any complex
algebraic manifold is based on a connection between the space-time
description and the quantum-mechanical principle of superposition. This
structure can be made consistent with the scheme of the canonical
quantization.

In the given paper we present new types of (super)algebras which are related
to the (super)twistor description. Usually, (super)twistors are considered
as objects of a complex projective (super)space with a given action of a
(super)conformal group. Nevertheless, basic twistor equation is invariant
not only with respect to (super)conformal group but also with respect to
so-named twistor shifts which affect the space-time coordinates. Extended
with such shifts (super)conformal group will have the transitive action on
spaces of flags.

We built straight ''extensions'' of the algebra of (super)conformal group
through the addition of pairs of ''twistor-like'' generators. This approach
is different from the tradition one, wherein the twistor manifolds are
described by the cosets $SL\left( 4,C\right) /P$, where $P$ is some
parabolic subgroup of $SL\left( 4,C\right) $ (see for example ref.\cite{Howe}
). From our point of view the approach presented in this paper allows to
find properties any twistor space more easily than the other known methods.

\section{Constructions of the "extended" (super)algebras}

\subsection{Background and short description}

\noindent
Firstly, we present a number of common knowledges related to homogeneous
spaces and to nonlinear realizations of groups.

Let ${\cal M}$ be a smooth manifold and $G$ is some Lie group having
differentiable action on ${\cal M}$, $G:{\cal M}\rightarrow {\cal M}$.
Subgroup $H_p\subset G$ is called a stationary subgroup of a point $p\in
{\cal M}$ iff $H_p:p=p$. If group $G$ acts transitively on ${\cal M}$, i.e. $%
{\cal M}$ is a homogeneous manifold, then there is a projection $\rho
(G/H)\rightarrow {\cal M}$, where $f\ni G/H,~h\ni H$ and $f\sim fh$. When
the structure of a smooth manifold can be introduced in $G/H$ and $\rho $
becomes a diffeomorphism. One say that manifold ${\cal M}$ is written in
Klein form ${\cal M}=G/H$.

Let us consider the homogeneous space (coset) $F=G/H$. The group $G$ is the
full isometry group of $F$, and $H$ is isotropy subgroup leaving the origin
invariant. The coordinates in $F$ are parameterized by group element $G\ni
g=g(\phi _1K_1,\phi _2K_2,\ldots )$, where $K_i$ denote the generators that
are not in $H$. The manifolds $G/H$ can also be thought of as sections of
fiber bundles ${\cal F}$, with total space $G$ and fiber $H$. These sections
are parameterized by the group elements $f\in F$.

The main idea of nonlinear realizations stems from the fact that the
isometries in $G$ that are not in $H$ are realized nonlinear on the fields $%
\phi _i$ in contrast the isometries in $H\subset G$. In this case one say
that the symmetry is broken from $G$ to $H$. It should be noted that the
action of any subgroup $S\subset G$ on $F$ will not be transitive. In
general case we will have
\begin{equation}
\label{orb}F=\bigcup\limits_k{\cal O}_k,
\end{equation}
where ${\cal O}_k$ are orbits of the subgroup $S$. That also takes place
when $S=H$.

It turns out that described above construction of the homogeneous spaces can
be directly applied to the (super)twistors with some additional proposes. We
will consider some minimal extensions of the (super)conformal group which
will be denoted by $G^{(\frac n2)}$, where $n$ will mark quantity of
additional pairs of so-named ''twistor-like'' generators $(q_\alpha ,s_{\dot
\alpha })$. From one hand, any function on coset $G^{(\frac n2)}/H$, where $%
H $ is (super)conformal group, provides the representation%
\footnote{totally reducible in general case} of $H$. But these functions
will not be functions on the (super)twistor space in the ordinary sense
because there is not the (super)twistor correspondence. From the other hand,
we can get it through the consideration both of the coset $G^{(\frac
n2)}/\tilde H$, where $\tilde H\subset H$, and some invariant hypersurfaces $%
\tilde T \subset G^{(\frac n2)}/\tilde H$ in this coset. The coset $%
G^{(\frac n2)}/\tilde H$ will play the role so-called the correspondence
space. We try to illustrate this thought some concrete examples below.

\subsection{"Extended" algebra for the twistors}

\noindent
Let us present some examples. Usual conformal algebra $c(1,3)\sim su(2,2)$
with the commutation relation
\begin{equation}
\label{fir}
\begin{array}{l}
{}~[P_{\alpha \dot \alpha },~K_{\beta \dot \beta }]=2\varepsilon_{\alpha \beta
}L_{\dot \alpha \dot \beta }+2\varepsilon_{\dot \alpha \dot \beta }L_{\alpha
\beta }+4i\varepsilon_{\alpha \beta }\varepsilon_{\dot \alpha \dot \beta }D,
\\
[1mm] ~[L_{\alpha \beta },~L_{\gamma \delta }]=\varepsilon_{\gamma (\beta
}L_{\alpha )\delta }+\varepsilon_{\delta (\beta }L_{\gamma )\alpha }, \\
[1mm] ~[L_{\dot \alpha \dot \beta },~L_{\dot \gamma \dot \delta
}]=\varepsilon_{\dot \gamma (\dot \beta }L_{\dot \alpha )\dot \delta
}+\varepsilon_{\dot \delta (\dot \beta }L_{\dot \gamma )\dot \alpha }, \\
[3mm]
\begin{array}{ll}
[D,~P_{\alpha \dot \alpha }]=-iP_{\alpha \dot \alpha }, & [D,~K_{\alpha \dot
\alpha }]=iK_{\alpha \dot \alpha }, \\
[1mm] [L_{\alpha \beta },~P_{\gamma \dot \gamma }]=\varepsilon_{\gamma
(\beta }P_{\alpha )\dot \gamma }, & and~~(\alpha \beta \rightarrow \dot
\alpha \dot\beta ), \\
[1mm] [L_{\alpha \beta },~K_{\gamma \dot \gamma }]=\varepsilon_{\gamma
(\beta }K_{\alpha )\dot \gamma }, & and~~(\alpha \beta \rightarrow \dot
\alpha \dot \beta ).
\end{array}
\end{array}
\end{equation}
where $L,~P,~D,~K$ are Lorentz, momentum, dilation and special conformal
generators respectively, we extend by one pair of generators $(q_\alpha
,s_{\dot \alpha })$ with following nontrivial commutators
\begin{equation}
\begin{array}{ll}
[P_{\alpha \dot \alpha },~s_{\dot \beta }]=2\varepsilon_{\dot \alpha \dot
\beta }q_\alpha , & [K_{\alpha \dot \alpha },~q_\beta
]=-2\varepsilon_{\alpha \beta }s_{\dot \alpha }, \\
[1mm][L_{\dot \alpha \dot \beta },~s_{\dot \gamma }]=\varepsilon_{\dot
\gamma (\dot \beta }s_{\dot \alpha )}, & [L_{\alpha \beta },~q_\gamma
]=\varepsilon_{\gamma (\beta }q_{\alpha )}, \\
[1mm][D,~q_\alpha ]=-\frac i2q_\alpha , & [D,~s_{\dot \alpha }]=\frac
i2s_{\dot \alpha }.
\end{array}
\end{equation}
As a result we will have $G^{(\frac 12)}$ ''extended'' algebra (which allows
us to describe the twistor $Z$, but not $\bar Z$). This is a minimal
extension of the conformal group. So, the conformal group action on the
coset $G^{(\frac 12)}/C(1,3)$ will be irreducible.

Now, we will illustrate how the equation, defining $\alpha $ -- plane, can
be produced from the outlined algebra. To get the twistor correspondence,
let us define another coset $F$ by the following choice%
\footnote{any other choice is also available, that will be
represent redefinition of the group coordinates}
\begin{equation}
F\ni f=\exp (\frac i2P_{\alpha \dot \alpha }x^{\dot \alpha \alpha })\exp
(iq_\alpha \omega ^\alpha +is_{\dot \alpha }\pi ^{\dot \alpha }).
\end{equation}
Here, we additionally introduced momentum generator in order to associate
parameters $(\omega ,~\pi )$ with Minkowsky space coordinates $x\in M^4$
which can be considered as complex-valued. Particularly, the conformal group
$C(1,3)$ will act nonlinearly on the parameters $x^{\dot \alpha \alpha }$.
The other parameters $(\omega ,~\pi )$ will transform linearly under the
conformal group action. From the latest follows that the correspondence will
depend on the choice of the initial point in $M^{4}$. To extract in $F$
subspaces that invariant under shifts of the twistor coordinates $(\omega
,~\pi )$ we consider left-invariant Cartan's form on $F$. One, being
restricted on $F$, is
\begin{equation}
f^{-1}df|_F=iq_\alpha (d\omega ^\alpha +idx^{\dot \alpha \alpha }\pi _{\dot
\alpha })+is_{\dot \alpha }d\pi^{\dot \alpha }+iP_adx^a.
\end{equation}
It is easy to see that the conditions, extracting the subspaces,
\begin{equation}
\label{basic} d\omega ^\alpha +idx^{\dot \alpha \alpha }\pi _{\dot \alpha
}=0,\quad d\pi _{\dot \alpha }=0
\end{equation}
completely determine $\alpha $ -- plane in proposition that $\omega =\omega
(x),~\pi =\pi (x)$. The conditions (\ref{basic}) can be also rewritten in
the form
\begin{equation}
\label{cond} dx^a\nabla _a\omega ^\alpha =0,\qquad dx^a\nabla _a\pi _{\dot
\alpha }=0,
\end{equation}
where covariant derivative $\nabla _{\alpha \dot \alpha }= {\frac \partial
{\partial x^{\alpha \dot \alpha }}}+i\pi _{\dot \alpha }{\frac \partial
{\partial \omega ^\alpha }}$ was defined. One is covariant derivative in the
{\em flat} twistor space. Now, parameters $\omega (x),~\pi (x)$ can be
considered as the usual twistor coordinates $Z^A=Z^A(\omega ,~\pi )$ on the
twistor space $T$ with the additional conditions
\begin{equation}
\label{twis}\nabla _{\alpha \dot \alpha }Z^A=0.
\end{equation}
These conditions give us an additional fibering of the coset space $F$, when
each fiber is $\alpha $ -- plane. The conditions (\ref{twis}) could be also
thought as identities on any twistor space. Then every function $\Phi (Z(x))$
on a twistor space has to obey the equation $\nabla _{\alpha \dot \alpha
}\Phi (Z)=0$. So, we find the definition of global twistors and {\em flat}
twistor spaces $T$.

In the same manner, the basic equations \cite{Penr, Merk}, that define local
twistor properties associated with conformal-flat space-time, arise on the
coset $F^{\prime }=G^{(\frac 12)}/SL(2,C)$. With choice coordinates on $%
F^{\prime }$ via
\begin{equation}
F^{\prime }\ni f^{\prime }=\exp (\frac i2x^{\dot \alpha \alpha }P_{\dot
\alpha \alpha })\exp (\frac i2\gamma ^{\dot \alpha \alpha }K_{\dot \alpha
\alpha })\exp (i\sigma D)\exp (i\omega ^\alpha q_\alpha +i\pi ^{\dot \alpha
}s_{\dot \alpha })
\end{equation}
conditions, analogous to (\ref{basic}), are
\begin{equation}
\tilde \nabla _{\delta \dot \delta }Z={\cal D}_{\delta \dot \delta }\left(
\begin{array}{c}
\omega ^\alpha \\
\pi _{\dot \alpha }
\end{array}
\right) +i\left(
\begin{array}{cc}
0 & \delta _\delta ^\alpha \delta _{\dot \delta }^{\dot \beta } \\
{\cal P}_{\delta \dot \delta ,\beta \dot \alpha } & 0
\end{array}
\right) \left(
\begin{array}{c}
\omega ^\beta \\
\pi _{\dot \beta }
\end{array}
\right) =0,
\end{equation}
where
$$
{\cal P}_{\delta \dot \delta ,\alpha \dot \alpha }=e^{2\sigma }(\partial
_{\delta \dot \delta }\gamma _{\alpha \dot \alpha }+2\gamma _{\delta \dot
\alpha }\gamma _{\alpha \dot \delta }),\quad {\cal D}_{\alpha \dot \alpha
}=e^\sigma \left( \nabla _{\alpha \dot \alpha }+\Gamma _{\alpha \dot \alpha
}\right) ,
$$
and $\Gamma _{\alpha \dot \alpha }$ are Lorentz connection in an appropriate
representation, $\nabla _{\alpha \dot \alpha }$ is the same as in (\ref{cond}%
).

{}From Cartan's equation
\begin{equation}
0\equiv d\Omega -\Omega \wedge \Omega =\left( P_a\tilde T_{~bc}^a+K_a\tilde
B_{~bc}^a+L_{ad}\tilde R_{~~bc}^{ad}\right) dx^b\wedge dx^c,
\end{equation}
where $\Omega =f^{\prime -1}df^{\prime }$, particularly follows that the
curvature of a conformal-flat space-time is defined by the {\em twistor}
connection ${\cal P}_a^b$ as $R_{dc}^{ab}=\delta _{[d}^{[a}{\cal P}
_{c]}^{b]} $ $\Leftrightarrow $ ${\cal P}_a^b=R_a^b-\frac 16R\delta _a^b$.

The equation, defining $\beta $ -- plane, can be easily found in the context
of nonlinear realizations of group $G^{(1)}$, which can be gained by adding
one more pair of generators $(q_{\dot \alpha },~s_\alpha )$ with appropriate
commutation relations. It is easy to check that the factor space $G^{(1)}/H$
will describe twistors $Z^A$, $\bar Z^A$ and the twistor space $T
\cup \bar T$. In order to get the standard dual relation between
$T$ and $\bar T$ , it is obvious to use Cartan -- Killing metric
of $G^{(1)}$ group projected into the coset $F$.

Further, we will present $G^{(2)}$ extended group that related to the
twistor sets $Z^A,~W^B$ and $\bar Z^A,~\bar W^B$. As it was demonstrated by
Penrose \cite{Penr} they make possible to describe massive fields and
particles.

Let us introduce the following set of ''twistor-like'' generators
$$
(q_{\alpha i},~q_{\dot \alpha i}),~(s_{\alpha i},s_{\dot \alpha i}),
$$
where $i=1,~2$. It was an essential point that in addition to the usual
conformal symmetry, from Jacoby identities we will have the ''helicity
charge'' $A$ and $SU(2)$ automorphism group
\begin{equation}
\label{two}
\begin{array}{l}
[J_{ij},q_l]=
\frac{1}{2}\varepsilon_{l(i}q_{j)}, \\ [1mm] [A,q_{\alpha i}]=iq_{\alpha i},
\\
[1mm] [A,q_{\dot \alpha }]=-iq_{\dot \alpha }.
\end{array}
\end{equation}
Their commutation relations with the generators of the conformal group are
\begin{equation}
\label{thr}
\begin{array}{l}
\begin{array}{ll}
[q_{\alpha i},~q_{\dot \alpha j}]=-\varepsilon_{ij}P_{\alpha \dot \alpha
},\quad & [s_{\alpha i},~s_{\dot \alpha j}]=\varepsilon_{ij}K_{\dot \alpha
\alpha }, \\
[1mm] [P_{\alpha \dot \alpha },~s_{\beta i}]=2\varepsilon_{\alpha \beta
}q_{\dot \alpha i}, & [K_{\dot \alpha \alpha },~q_{\beta
i}]=-2\varepsilon_{\alpha \beta }s_{\dot \alpha i}, \\
[1mm] [P_{\alpha \dot \alpha },~s_{\dot \beta i}]=2\varepsilon_{\dot \alpha
\dot \beta }q_{\alpha i}, & [K_{\alpha \dot \alpha },~q_{\dot \beta
i}]=-2\varepsilon_{\dot \alpha \dot \beta }s_{\alpha i},
\end{array}
\\
[7mm] ~[q_{\alpha i},~s_{\beta i}]=\varepsilon_{\alpha \beta
}\varepsilon_{ij}(iD-
{\frac{3i}2}A)+2\varepsilon_{\alpha \beta }J_{ij}+\varepsilon_{ij}L_{\dot
\alpha \dot \beta }, \\ [1mm] ~[q_{\dot \alpha i},~s_{\dot \beta
i}]=\varepsilon_{\dot \alpha \dot \beta }\varepsilon_{ij}(iD+{\frac{3i}2}
A)+2\varepsilon_{\dot \alpha \dot \beta }J_{ij}+\varepsilon_{ij}L_{\dot
\alpha \dot \beta }.
\end{array}
\end{equation}
The full set of commutators defining algebra of $G^{(2)}$ group consist from
(\ref{fir}), (\ref{two}) and (\ref{thr}).

It can be seen that subalgebra $G^{(1)}\subset G^{(2)}$ is associated with
generators $(q_{\alpha 1},~q_{\dot \alpha 2}),$ $(s_{\alpha 2},s_{\dot
\alpha 1}).$ Let us consider this subalgebra some more detail. We restrict
our attention to some affirmations in the light of which the construction of
global twistors correspondence occurs amazingly simple. Obviously, that
twistors of the fundamental $SU(2,2)$ group representation having all known
properties can be found in the context of nonlinear realizations of $G^{(1)}$
from trivially contracted target algebra.

We fix the coset through the following expression:
\begin{equation}
\exp (\frac i2P^{\dot \alpha \alpha }x_{\dot \alpha \alpha })\exp (\frac{2i}
5a\tilde A)\exp (i{\omega }^\alpha q_{\alpha 1}+i\pi ^{\dot \alpha }s_{\dot
\alpha 1})\exp (i\omega ^{\dot \alpha }q_{\dot \alpha 2}+i\pi ^\alpha
s_{\alpha 2}),
\end{equation}
where $\tilde A=A+2J_{12}$. On the intersection of $\alpha $ and $\beta $
plains we have
\begin{equation}
\omega ^{\dot \alpha }\pi _{\dot \alpha }-\omega ^\alpha \pi _\alpha =0,
\end{equation}
which can be fulfilled then
$$
\omega ^\alpha \pi _\beta -\pi ^\alpha \omega _\beta =\delta _{~\beta
}^\alpha ,\quad (\alpha ,\beta \rightarrow \dot \alpha ,\dot \beta ).
$$
{\bf Notice:} {\em It is easy to see the analogy with the harmonic space
construction {\rm \cite{Galp}}, where $(\omega,~ \pi)$ play the role of
harmonical coordinates on the sphere $S^{2} = SU(2)/U(1).$ The connection
between the harmonic space construction and the algebra $G^{(1)}$ is
available by the automorphysm $SU(2) \subset G^{(1)}$ subgroup.}

As this take place, real Minkowsky space coordinates can be completely
''gauged away'' by
$$
x^{\alpha \dot \alpha }=-i\omega ^\alpha \omega ^{\dot \alpha }.
$$
Conditions defining functions $\Phi (x,\omega ,\bar \omega )$ on the twistor
space $T\cap \bar T$ are two constrains $\nabla _{(a)}\Phi =0,$ $\nabla
_{\alpha \dot \alpha }\Phi =0$. The solutions of the first constrain
\begin{equation}
\label{fcond} \nabla _{(a)}\Phi =({\frac \partial {\partial a}}-i\omega
^\alpha {\frac \partial {\partial \omega ^\alpha }}+i\omega ^{\dot \alpha }
\frac{\partial}{\partial\omega^{\dot{\alpha}}})\Phi =0,
\end{equation}
give us left or right helicity fields $\psi (x)$
\begin{equation}
\begin{array}{ll}
n=0,~1,~2,\ldots & \leadsto ~~\Phi =e^{ina}\omega ^{\alpha _1}\cdots \omega
^{\alpha _n}\psi _{\alpha _1\cdots \alpha _n}(x), \\
[6mm] n=-1,~-2,\ldots & \leadsto ~~\Phi =e^{ina}\omega ^{\dot
\alpha_1}\cdots \omega ^{\dot \alpha _n}\psi _{\dot \alpha _1\cdots \dot
\alpha _n}(x),
\end{array}
\end{equation}
where $n$ parameterize the solutions of (\ref{fcond}) and define homogeneous
degree of the functions $\Phi (Z)$ or $\Phi (\bar Z)$. The second constrain
\begin{equation}
\label{scond} \nabla _{\alpha \dot \alpha }\Phi =(\frac{\partial}{\partial
x^{\alpha \dot{\alpha}}}-\pi _{\dot{\alpha}}\frac{\partial}
{\partial\omega^{\alpha}}-\pi _\alpha \frac{\partial} {\partial \omega
^{\dot \alpha }})\Phi =0
\end{equation}
is the free equation of motion massless fields. Essentially, that on the
subspace $T\cap \bar T$, either left or right
helicity fields $\psi _\alpha \ldots $ can ''live''. It seems that the
constrains (\ref{fcond},~\ref{scond}) are agreed upon conditions defined
representations of the conformal group. Outlined above method can be also
applied for constructing a large number of cosets $G^{(2)}/H$ by looking at
the parabolic subgroups of $G^{(2)}$.

\subsection{"Extended" algebra for the supertwistors}

\noindent
The supertwistor correspondence can be also described in the context of
coset realizations of ''extended'' supergroups $SG^{(n|N)}$, where $N$
denotes quantity of fermionic spinoral generators in associated
superconformal group. In this section we will describe $SG^{(2|1)}$
supergroup. The algebra of the group $SG^{(2|1)}$ is obtained from relation (%
\ref{fir}, \ref{two}, \ref{thr}) and $N=1$ conformal supersymmetry
\begin{equation}
\label{for}
\begin{array}{ll}
\{Q_\alpha ,Q_{\dot{\alpha}}\}=P_{\alpha\dot{\alpha}}, & \{R_\alpha ,R_{
\dot{\alpha}}\}=K_{\alpha \dot \alpha }, \\ [1mm] [P_{\alpha \dot{\alpha}
},R_{\beta} ]=2i\varepsilon_{\alpha \beta }Q_{\dot{\alpha}}, & [K_{\alpha
\dot{\alpha}},Q_{\beta}]=2i\varepsilon_{\alpha \beta }R_{\dot{\beta}}, \\
[1mm] [P_{\dot{\alpha} \alpha },R_{\dot{\beta}}]=2i\varepsilon_{\dot{\alpha}
\dot{\beta}}Q_{\alpha} , & [K_{\alpha \dot{\alpha}},R_{\dot{\beta}}]=
2i\varepsilon_{\dot{\alpha}\dot{\beta}}R_{\alpha},
\end{array}
\end{equation}
by adding $(Q_i,~R_i)$ generators and charge $B$ having odd Grassmannian
parities
\begin{equation}
\label{fiv}
\begin{array}{ll}
[B,~Q_{\alpha (\dot{\alpha})}]=\pm \frac{1}{2}Q_{\alpha (\dot{\alpha})}, &
[B,~Q_i]=
\frac{1}{2}Q_i, \\ [1mm] [B,~R_{\alpha (\dot \alpha )}]=\mp \frac{1}{2}
R_{\alpha (\dot \alpha )}, & [B,~R_i]=-
\frac{1}{2}R_i, \\ [1mm] [A,~Q_i]=-iQ_i, & [A,~R_i]=iR_i.
\end{array}
\end{equation}
Jacoby identities fix the other relations
\begin{equation}
\label{six}
\begin{array}{l}
\begin{array}{l}
[q_{\alpha i},~s_{\beta j}]=\varepsilon_{ij}\varepsilon_{\alpha \beta
}(iD+B-3i/2A)+\varepsilon_{ij}L_{\alpha \beta }+2\varepsilon_{\alpha \beta
}J_{ij}, \\
[1mm] [q_{
\dot{\alpha} i},~s_{\dot{\beta} j}]=\varepsilon_{ij} \varepsilon_{\dot{%
\alpha }\dot {\beta}}(iD-B+3i/2A)+\varepsilon_{ij}L_{\dot{\alpha}\dot{\beta}
}+ 2\varepsilon_{\dot{\alpha}\dot{\beta}}J_{ij}, \\ [1mm]
\{Q_i,~R_j\}=\varepsilon_{ij}(2iB+
\frac{1}{2}A)-iJ_{ij}, \\ [1mm] \{Q_{
\dot{\alpha}}R_{\dot{\beta}}\}=\varepsilon_{\dot{\alpha}\dot{\beta} }(D-3iB+
\frac{1}{2}A)-iL_{\dot{\alpha}\dot{\beta}}, \\ [1mm] \{Q_\alpha
R_\beta\}=\varepsilon_{\alpha \beta }(D+3iB-
\frac{1}{2}A)-iL_{\alpha \beta }, \\ [1mm]
\end{array}
\\
[6mm]
\begin{array}{ll}
\{Q_i,~Q_{\dot{\alpha}}\}=q_{\dot{\alpha} i}, & \{R_i,~R_{
\dot{\alpha}}\}=s_{\dot{\alpha} i}, \\ [1mm] \{R_i,~Q_{\alpha}
\}=-iq_{\alpha i}, & \{Q_i,~R_\alpha \}=is_{\alpha i}, \\
[1mm] [q_{\delta i},~Q_j]=-\varepsilon_{ij}Q_\delta , & [s_{\alpha
i},~R_j]=\varepsilon_{ij}R_{\alpha} , \\
[1mm] [q_{\alpha i},~R_{\beta}]=-2\varepsilon_{\alpha \beta }R_i, &
[s_{\alpha i},~Q_\beta ]=2\varepsilon_{\alpha \beta }Q_i, \\
[1mm] [q_{\dot{\alpha}i},~R_{\dot{\beta} }]=2i\varepsilon_{\dot{\alpha}\dot{
\beta}}Q_i, & [s_{
\dot{\alpha} i},~Q_{\dot{\beta}}]=2i\varepsilon_{\dot{\alpha}\dot{\beta}
}R_i, \\ [1mm] [q_{\dot{\alpha} i},~R_j]=i\varepsilon_{ij}Q_{\dot{\alpha} },
& [s_{\dot{\alpha} i},~Q_j]=i\varepsilon_{ij}R_{\dot{\alpha} }.
\end{array}
\end{array}
\end{equation}
Various versions of the equations defining left(right) and central $\alpha $(%
$\beta $) -- superplanes and the properties of both global and local
supertwistors are easily determined. So, $\alpha $ -- superplanes are
defined by the following condition
\begin{equation}
d\omega^{\alpha} =dx^{\dot{\alpha}\alpha }\pi _{\dot{\alpha} }+2d\theta
^\alpha \xi ,\quad d\xi =2d\bar{\theta}^{\dot{\alpha} }\pi _{\dot{\alpha}},
\end{equation}
where parameters $(\omega ,\pi ,\xi )$ defining supertwistor components are
associated with generators $(q_{2\alpha },s_{2\dot{\alpha}},R_2)$.

Because there is a formal generalization of the contour integrals method for
constructing of chiral superfields, it seems that there is a local
equivalence between the general conformal supermultipletes and their
potentials. One may suppose, that there is some cohomological description
including explicit expressions for superfields and prepotential
representations.

\section{Summary and further perspectives}

\noindent
In this paper we proposed some class of the ''twistor-like'' extensions $%
G^{(n|N)}$ for $n=1,2;~N=0,1$ (super)conformal group and illustrated how the
variants of the (super)twistor correspondence can arise on the cosets $%
G^{(n|N)}/H$. This approach seems us promising for more detailed analysis
twistor field models. Particularly, it can be useful for the models given
below.

It is well known that the obstructions of integrability of some nonlinear
field models (such as self-dual gravity, YM) in the twistor approach vanish.
The general solutions of these models are defined by the intrinsic
properties of complex manifolds rather then by the equations of motion or by
action principles. An interaction therewith is encoded in deformations of
the complex structure of twistor space. The description of appropriate
deformations, which are connected with whole Poincar\'e and (super)conformal
gravity, is still an open problem.

Proposed in ref. \cite{Hodg} radical modification of the standard string
theory to a four-dimensional field theory, where Riemann surfaces and
holomorphic functions of 2D CFT are replaced by generalized twistor spaces
and holomorphic ''one-functions'' respectively, seems to be to have a set of
remarkable properties of exactly solvable as it takes place in 2D.

A more detailed consideration of the mentioned problems will be done in
subsequent publications.

\end{document}